\documentclass[aps,pre,preprint, showpacs, superscriptaddress]{revtex4}
\usepackage[dvips]{graphicx}
\usepackage{latexsym}
\usepackage{natbib}

\begin{document}
\title{Conservative model for synchronization problems in complex networks.}
\author{C.~E.~La~Rocca}
\affiliation{Instituto de Investigaciones F\'isicas de Mar del Plata (IFIMAR)-Departamento de F\'isica, Facultad de Ciencias Exactas y Naturales, Universidad Nacional de Mar del Plata-CONICET, Funes 3350, (7600) Mar del Plata, Argentina.}
\author{L. A. Braunstein}
\affiliation{Instituto de Investigaciones F\'isicas de Mar del Plata (IFIMAR)-Departamento de F\'isica, Facultad de Ciencias Exactas y Naturales, Universidad Nacional de Mar del Plata-CONICET, Funes 3350, (7600) Mar del Plata, Argentina.}
\affiliation{Center for Polymer Studies, Boston University, Boston, Massachusetts 02215, USA}
\author{P. A. Macri}
\affiliation{Instituto de Investigaciones F\'isicas de Mar del Plata (IFIMAR)-Departamento de F\'isica, Facultad de Ciencias Exactas y Naturales, Universidad Nacional de Mar del Plata-CONICET, Funes 3350, (7600) Mar del Plata, Argentina.}
\begin{abstract}
In this paper we study the scaling behavior of the interface fluctuations
(roughness) for a discrete model with conservative noise on complex
networks. Conservative noise is a noise which has no external flux of
deposition on the surface and the whole process is due to the diffusion. It
was found that in Euclidean lattices the roughness of the steady state $W_s$
does not depend on the system size. Here, we find that for Scale-Free
networks of $N$ nodes, characterized by a degree distribution $P(k)\sim
k^{-\lambda}$, $W_s$ is independent of $N$ for any $\lambda$. This behavior
is very different than the one found by Pastore y Piontti {\it et. al}
[Phys. Rev. E {\bf 76}, 046117 (2007)] for a discrete model with
non-conservative noise, that implies an external flux, where $W_s \sim \ln N$
for $\lambda < 3$, and was explained by non-linear terms in the analytical
evolution equation for the interface [La Rocca {\it et. al}, Phys. Rev. E
{\bf 77}, 046120 (2008)]. In this work we show that in this processes with
conservative noise the non-linear terms are not relevant to describe the
scaling behavior of $W_s$.
\end{abstract}


\maketitle

\section{Introduction}

It is known that many physical and dynamical processes employ complex
networks as the underlying a substrate. For this reason many studies on
complex networks are focused not only in their topology but also in the
dynamic processes that run over them. Some examples of these kind of
dynamical processes on complex networks are cascading failures
\cite{Motter_prl}, traffic flow \cite{Lopez_transport,zhenhua}, epidemic
spreadings \cite{Pastorras_PRL_2001} and synchronization
\cite{Jost-prl,Korniss07}. In particular, synchronization problems are very
important in the dynamics and fluctuations in task completion landscapes in
causally constrained queuing networks \cite{Kozma05}, in supply-chain
networks based on electronic transactions \cite{Nagurney}, brain networks
\cite{JWScanell}, and networks of coupled populations in the synchronization
of epidemic outbreaks \cite{eubank_2004}.  For example, in the problem of
the load balance on parallel processors the load is distributed between the processors. If the system is
not synchronized, few processors have low load and they will have to wait for
the most loaded processors to finish the task. The nodes (processors) of the system
 have to synchronize
with their neighbors to ensure causality on the dynamics. The computational
time will be given by the most loaded processors, thus synchronizing the
system is equivalent to reduce or optimize the computational
time. Synchronization problems deal with the optimization of the fluctuations
of some scalar field $h$ (load in processing) in the system that will be
optimal synchronized minimizing those fluctuations. To analyze
synchronization problems is customary to study the height fluctuations of a
non-equilibrium surface growth. If the scalar field on the nodes represents
the interface height at each node, its fluctuations are characterized by the
average roughness $W(t)$ of the interface at time $t$, given by $W \equiv
W(t) = \left\{1/N \sum_{i=1}^N (h_i-\langle h \rangle)^2\right\}^{1/2},$
where $h_i \equiv h_i(t)$ is the height of node $i$ at time $t$, $ \langle h
\rangle$ is the mean value on the network, $N$ is the system size, and $\{
. \}$ denotes an average over configurations. Pastore y Piontti {\it et. al}
\cite{anita} studied this mapping in Scale-Free (SF) networks
\cite{Barabasi_sf} of broadness $\lambda$ and size $N$ using a surface
relaxation growth model (SRM) \cite{family} with non-conservative noise and
found that for $\lambda <3$ the saturation roughness $W_s$ scales as $W_s
\sim \ln N$. Later, the evolution equation for the interface in this model
was derived analytically \cite{crispre} for any complex networks.
The derived evolution equation has non-linear terms as a consequence of the heterogeneity of the
network that together with the non-conservative noise are necessary to explain the $W_s
\sim \ln N$ behavior for $\lambda <3$. However, there exist many physical
processes where the noise is conservative and cannot be modeled as a flux
deposition on a surface. In models without external flux, where particles
are moved by diffusion, the total volume of the system remains
unchanged. Examples of this kind of process are thermal fluctuations,
diffusion by an external agent such as an electric field, load balance of
parallel processors where the total load in the system is constant over a
certain time interval. For the last example, the only flux is due to
diffusion of the load from a processor to another. Though not extensively,
conservative noise has already been studied in Euclidean lattices
\cite{cneucli1,cneucli2} and it was found that $W_s$ does not depend on the
system size $L$. The evolution equation of this  process in Euclidean lattices can be well represented
by an Edwards-Wilkinson (EW) process \cite{EW} with conservative noise.
In this paper we study this model in SF networks  by simulations
of the discrete model (Section \ref{modelo}) and derive analytically its
evolution equation for any complex networks (Section \ref{ecuacion}). 
Those networks represents better the heterogeneity
in the contacts in real systems, like the Internet, the WWW, networks of routers, etc..
We applied the mean-field approximation to the evolution equation and show
that the scaling behavior of $W_s$ with $N$ (Section \ref{campomedio}) is only due to finite size effects.
To our knowledge this class of model was never studied before in complex networks.

\section{Simulations of the discrete model}\label{modelo}

In this model, at each time step a node $i$ is chosen with probability
$1/N$. If we denote by $v_i$ the nearest-neighbor nodes of $i$ , then (1) if
$h_i < h_j$, $\forall j \in v_i$ $\Rightarrow$ the scalar fields remains
unchanged, else (2) if $h_j < h_n$, $\forall n \not= j \in v_i$ $\Rightarrow
h_j = h_j+1$ and $h_i = h_i-1$. In that way the total height of the interface
is conserved and we have that the average height is constant. We
measure the roughness $W$ for SF networks, characterized by a power law tail
in the degree distribution $ P(k) \sim k^{-\lambda}$, where $k_{max} \ge k
\ge k_{min}$ is the degree of a node, $k_{max}$ is the maximum degree,
$k_{min}$ is the minimum degree and $\lambda$ measures the broadness of the
distribution \cite{Barabasi_sf}. 
To build the SF network we use the Molloy Reed (MR) \cite{Molloy} algorithm or configurational model.

\begin{figure}
\includegraphics[width=0.4\textwidth]{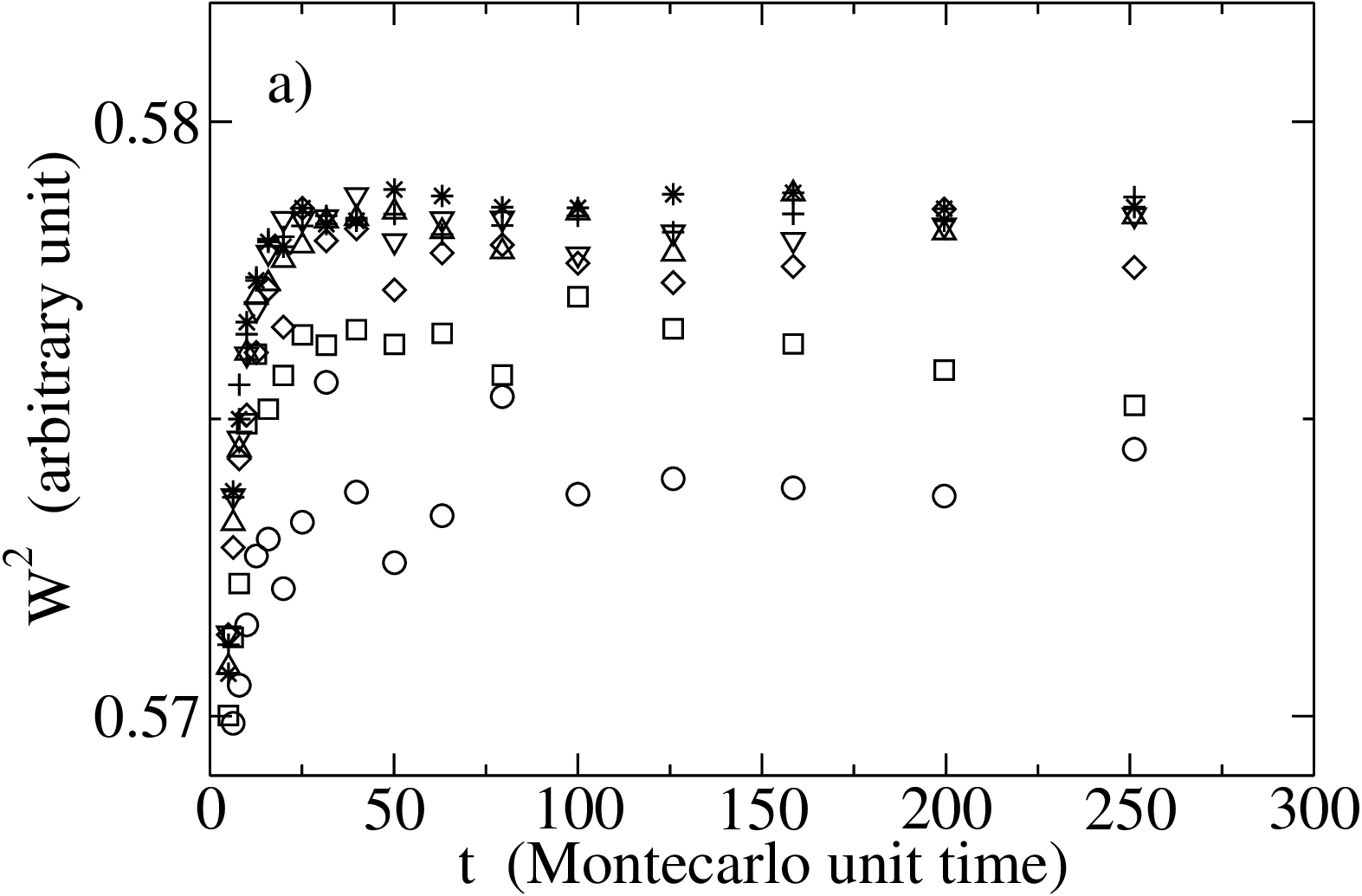}
\hspace{1cm}
\includegraphics[width=0.4\textwidth]{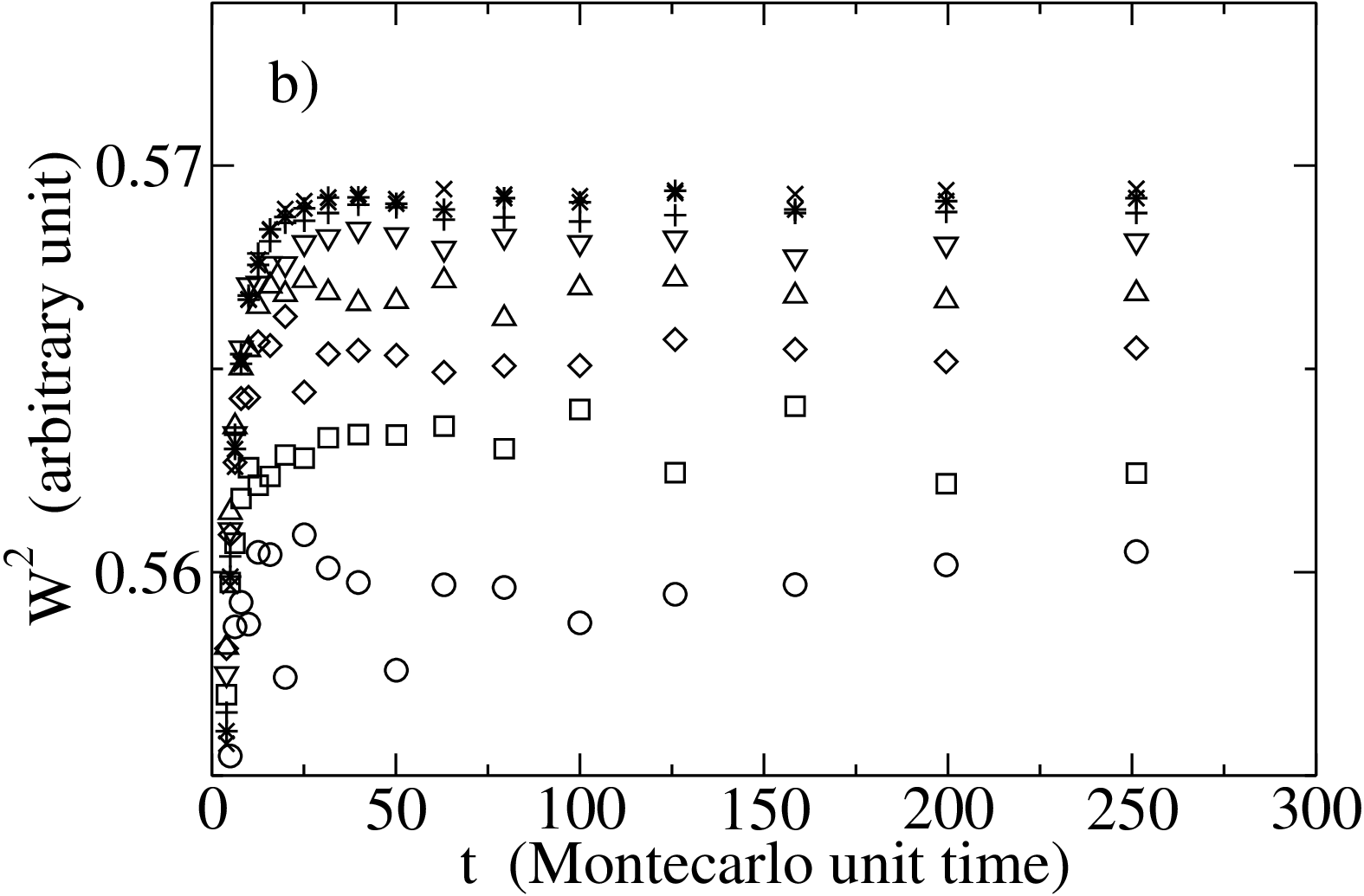}
\caption{$W^2$ as a function of $t$ for the discrete model for a) $\lambda=\
  3.5$ for $N=64$ ($\circ$), $128$ ($\Box$), $256$ ($\diamond$), $512$
  ($\bigtriangleup$), $1024$ ($\bigtriangledown$), $2048$ ($+$), $3072$
  ($\star$) and $4096$ ($X$) and b) $\lambda=\ 2.5$ for $N=64$ ($\circ$),
  $128$ ($\Box$), $256$ ($\diamond$), $512$ ($\bigtriangleup$), $768$
  ($\bigtriangledown$), $1024$ ($+$) and $1280$ ($\star$). Each curve was
  obtained with 10.000 realizations.\\\label{fig.1}}
\end{figure}

In Fig.~\ref{fig.1} we plot $W^2$ as a function of the time $t$ for different
system sizes and in Fig.~\ref{fig.2} the steady state $W_s^2$ as a function
of $N$, for (a) $\lambda =3.5$ and (b) $\lambda =2.5$. We can see that
$W_s^2$ increases with $N$ but, as we will show later, this dependence in the
system size is only due to finite size effects introduced by the correlated
nature (dissortative) of the MR algorithm \cite{boguna}.  For all the results
we use $k_{min}=2$ in order to ensure that the network is fully connected
\cite{cohen}, and assume that the initial configuration of $\{h_i\}$ is
randomly distributed in the interval $[-0.5,0.5]$. Then, we have that
$\langle h \rangle=0$.

\section{Derivation of the stochastic continuum equation}\label{ecuacion}

Next we derive the analytical evolution equation for the scalar field $h_i$
for every node $i$ in the conservative model in random graphs. The procedure
chosen here is based on a coarse-grained (CG) version of the discrete
Langevin equations obtained from a Kramers-Moyal expansion of the master
equation \cite{VK,Vveden,lidia}. The discrete Langevin equation for the
evolution of the height in any growth model is given by \cite{Vveden,lidia}
\begin{eqnarray}\label{eqh}
\frac{\partial h_i}{\partial t}= \frac{1}{\tau}K^1_i + \eta_i,
\end{eqnarray}
where $K^1_i$ takes into account the deterministic growth rules that produces
the evolution of the scalar field $h_i$ on node $i$, $\tau=N \delta t$ is the
mean time of attempts to change the scalar fields of the interface, and
$\eta_i$ is a noise with zero mean and covariance given by
\cite{Vveden,lidia}
\begin{equation}\label{ruido}
\{\eta_i(t)\eta_j(t')\}=\frac{1}{\tau}K_{ij}^2\delta(t-t')\ .
\end{equation}

More explicitly, $K^1_i$ and $K_{ij}^2$ are the two first moments of the
transition rate and they are given by
\begin{equation}\label{eqreglas}
K^1_i = \sum_{j=1}^{N} A_{ij}\left[P_{ij}-P_{ji}\right]\ ,
\end{equation}
\begin{equation}\label{momentodos}
K_{ij}^2=\frac{1}{\tau}K^1_i\delta_{ij}-\frac{1}{\tau}\sum_{n=1}^{N} A_{in}(P_{in}+P_{ni})(\delta_{nj}-\delta_{ij})\ ,
\end{equation}
where $\{A_{ij}\}$ is the adjacency matrix ($A_{ij}=1$ if $i$ and $j$ are
connected and zero otherwise) and $P_{ij}$ is the rule that represents the
growth contribution to node $i$ by relaxation from its neighbor $j$. In our
model the network is undirected, then $A_{ij}=A_{ji}$. As the rules for this
model are very complex if we allow degenerate scalar fields, we simplify the
problem taking random initial conditions [See discrete rules on Sec.\ref{modelo}]. Thus,
\[ P_{ij}= \Theta(h_{j}-h_{i})\prod_{n \in v_j}[1-\Theta(h_{i}-h_{n})]\ ,\]
where $\Theta$ is the Heaviside function given by $\Theta(x)=1$ if $x \geq 0$
and zero otherwise, with $x=h_t-h_s \equiv \Delta h$. Without lost of
generality, we take $\tau=1$.

In the CG version $\Delta h \rightarrow 0$; thus after expanding an analytical
representation of $\Theta(x)$ in Taylor series around $x=0$ to first order in
$x$, we obtain
\begin{eqnarray}\label{nonliner}
K^1_i&&=\ c_0\ \sum_{j=1}^{N}A_{ij}\ \left[\Omega(k_j)-\Omega(k_i)\right]\ +\
  c_1\ \sum_{j=1}^{N}A_{ij}\ \left[\Omega(k_j)+\Omega(k_i)\right]\
  (h_j-h_i)\nonumber\\ &&+\ \frac{c_1c_0}{(1-c_0)}\
  \sum_{j=1}^{N}A_{ij}\Omega(k_j)\
  \left[\sum_{n=1,n\not=i}^{N}A_{jn}(h_n-h_i)\right] + O((\Delta h)^2)\ ,
\end{eqnarray}
where $c_0$ and $c_1$ are the first two coefficients of the expansion of
$\Theta(x)$ and $\Omega(k_i)=(1-c_0)^{k_i-1}$ is the weight on the link $ij$
introduced by the dynamic process.

Notice that the network is undirected and the noise is conservative, thus the
average noise correlation [see Eq.~(\ref{ruido})] is $\langle \eta_i(t) \eta_j
(t^{'}) \rangle=0$, where $\langle ~~\rangle$ represents average over all the nodes of the
network. Notice that in Eq.~(\ref{nonliner}) the non-linear terms are
disregarded. As we will show below, for this conservative noise model these
terms are not necessary to explain the scaling behavior of $W_s$ with $N$.

We numerically integrate our evolution equation Eq.~(\ref{eqh}) in SF
networks using the Euler method with a representation of the Heaviside
function given by $\Theta(x)=(1+ \tanh[U(x+z)])/2$, where $U$ is the width
and $z= 1/2$ \cite{lidia}. With this representation, $c_0=(1+\tanh[U/2])/2$
and $c_1=(1-\tanh^2[U/2])\ U/2$. We assume that the initial configuration of
$\{h_i\}$ is randomly distributed in the interval $[-0.5,0.5]$ and for the
conservative noise we used the algorithm described in \cite{ruidoconser}: at
each time step, for every node in the network and for any of its nearest
neighbor we add a random number in the interval [-0.5,0.5] and remove this
amount to one of the nearest neighbor nodes.

In Fig.~\ref{fig.3} we plot $W^2$ as a function of $t$ from the integration
of Eq.~(\ref{eqh}) for (a) $\lambda=3.5$ and (b) $\lambda=2.5$, and different
values of $N$ with $k_{min}=2$. For the time step integration we chose
$\Delta t \ll 1/k_{max}$ according to Ref~\cite{pasointeg}. In
Fig.~\ref{fig.4} we plot the steady state $W_s^2$ as a function of $N$ for
(a) $\lambda=3.5$ and (b) $\lambda=2.5$. We can see that $W_s^2$ depends
weakly on $N$, but as shown below this size dependence is due to finite size
effects introduced by the MR construction. Next we derive the mean field
approximation in order to explain the nature of the corrections to the
scaling.

\section{Mean field approximation for the evolution equation}\label{campomedio}

We apply a mean field (MF) approximation to the linear terms of
Eq.~(\ref{nonliner}). In this approximation we consider $1 \ll k_{min} \ll
k_{max}$. Taking $C_{ij}=A_{ij}\Omega(k_j)$ and
$T_{ijn}=A_{ij}A_{jn}\Omega(k_j)$, then
\begin{eqnarray}\label{nonliner2}
K^1_i&&=\ c_0\ \left[C_i-k_i\Omega(k_i)\right]\ +\ c_1
  C_i\left[\sum_{j=1}^{N}\frac{C_{ij}h_j}{C_i}-h_i\right]\nonumber\\ &&+\ c_1
  \Omega(k_i)k_i\left[\sum_{j=1}^{N}\frac{A_{ij}h_j}{k_i}-h_i\right]\\ &&+\
  \frac{c_1c_0}{(1-c_0)}\ T_i
  \left[\sum_{j=1}^{N}\sum_{n=1,n\not=i}^{N}\frac{T_{ijn}h_n}{T_i}-h_i\right]\
  ,\nonumber
\end{eqnarray}
where
\begin{eqnarray}\label{peso1}
C_i=&\sum_{j=1}^{N}C_{ij}\ ;\nonumber\\
T_i=&\sum_{j=1}^{N}\sum_{n=1,n\not=i}^{N}T_{ijn}\ .
\end{eqnarray}
Disregarding the fluctuations, we take $ \sum_{j=1}^{N} A_{ij}h_j
/k_i \approx \langle h \rangle$, $ \sum_{j=1}^{N}C_{ij}h_j/C_i \approx
\langle h \rangle$, and $\sum_{j=1}^{N}\sum_{n=1,n\not=i}^{N}T_{ijn}h_n/T_i
\approx \langle h \rangle $. From Eq.~(\ref{peso1}), we can approximate $C_i$
by $C_i(k_i) \approx k_i\int_{k_{min}}^{k_{max}} P(k|k_i)\ \Omega(k)\ d k$
\cite{Korniss07}, where $P(k|k_i)$ is the conditional probability that a node
with degree $k_i$ is connected to another with degree $k$. For uncorrelated
networks $P(k|k_i) = k P(k) /\langle k \rangle$ \cite{Barabasi_sf}, then
$C_i(k_i) \approx I_1\ k_i/\langle k\rangle$ with
$I_1=\int_{k_{min}}^{k_{max}} P(k)\ k\ \Omega(k)\ dk$ .  Making the same
assumption for $T_i$, we obtain $T_i(k_i) \approx I_2\ k_i/\langle k\rangle$
with $I_2=\int_{k_{min}}^{k_{max}} P(k)\ k\ (k-1)\ \Omega(k)\ dk$. Then, the
linearized evolution equation for the heights in the MF approximation can be
written as
\begin{eqnarray}\label{eqaprox}
\frac{\partial{h_i}}{\partial t}&=& F_i(k_i)+ \nu_i(k_i)\ (\langle h
\rangle-h_i) + \eta_i\ ,
\end{eqnarray}
where $F_i(k_i)=c_0k_i\left[I_1/\langle k \rangle-\Omega(k_i)\right]$
represents a local driving force, $\nu_i(k_i)=c_1k_i(b+\Omega(k_i))$ is a
local superficial tension-like coefficient with $b=[I_1 +
I_2c_0/(1-c_0)]/\langle k \rangle$. This mean field approximation reveals the
network topology dependence through $P(k)$.

Taking the average over the network in Eq.~(\ref{eqaprox}), $\partial \langle
h \rangle/\partial t=1/N\sum_{i=1}^N F_i=0$, then $\langle
h \rangle$ is constant in time. The solution of
Eq.~(\ref{eqaprox}) \cite{VK} is given by
\begin{eqnarray}\label{solucion}
  h_i(t)&=& \int_{0}^{t}e^{-\nu_i(t-s)}\ (F_i+\nu_i\langle h \rangle+\eta_i(s))\ ds\nonumber \\
 &=& \frac{F_i+\nu_i\langle h \rangle}{\nu_i} (1-e^{-\nu_it}) + \int_{0}^{t}e^{-\nu_i(t-s)}\eta_i(s)\ ds\ .
\end{eqnarray}

Using Eq.~(\ref{solucion}) and the fact that in our model with the initial conditions we use $\langle h \rangle
=0$, we find the two-point correlation function
\begin{eqnarray*}
\left\{h_i(t_1) h_j(t_2)\right\}&&=\
  \left(\frac{F_i}{\nu_i}\right)\left(\frac{F_j}{\nu_j}\right)
  (1-e^{-\nu_it})(1-e^{-\nu_jt})\\ \nonumber +&&
  \int_{0}^{t_2}\int_{0}^{t_1}e^{-\nu_i(t_1-s_1)}e^{-\nu_j(t_2-s_2)}\left\{\eta_i(s_1)\eta_j(s_2)\right\}\
  ds_1ds_2\ .\nonumber
\end{eqnarray*}
For $t>max\left\{1/\nu_i \right\}$, we can write $W_s$ as
\begin{equation}\label{rugsat}
W_s^2=\{< h_i^2>\}=\frac{1}{N}\sum_{i=1}^N\left(\frac{F_i}{\nu_i}\right)^2
  + \frac{1}{N}\sum_{i=1}^N\frac{K_{ii}^2}{\nu_i}\ ,
\end{equation}
where $K_{ii}^2$ [See Eq.~(\ref{momentodos})] is given by
\[ K_{ii}^2=\sum_{j=1}^{N} A_{ij}\left[P_{ij}+P_{ji}\right] \approx c_0k_i\
\left[\frac{I_1}{\langle k \rangle}+\Omega(k_i)\right]\ .\]

For SF networks it can be shown that $I_1, I_2 \sim \mbox{const.}+k_{max}
\exp(-k_{max}\mbox{const.})$, where $k_{max}\sim N^{1/(\lambda -1)}$ for MR
networks; thus we can considerer the quantities $I_1$ and $I_2$ as
independent of $N$.

From Eq.~(\ref{rugsat}) and using the expressions for $F_i$, $\nu_i$ and
$K_{ii}^2$, we have
\begin{equation}\label{rugsatsuma}
W_s^2=\left(\frac{c_0}{c_1}\right)^2
  B^2\frac{1}{N}\sum_{i=1}^N\left(f_{-}(k_i)\right)^2 +\frac{c_0}{c_1}B
  \frac{1}{N}\sum_{i=1}^N f_{+}(k_i)\ ,
\end{equation}
where
\[ f_{\pm}(k_i)=\frac{1\pm \frac{\langle k \rangle}{I_1}\ \Omega(k_i)}{1 +\frac{\langle
k \rangle}{I_2}\frac{c_0}{1-c_0}\ \Omega(k_i)}\ ,\] and $B=1/(1 + c_0
I_2/(1-c_0)I_1)$. Taking the continuum limit we find another expression for
Eq.~(\ref{rugsatsuma}) as
\begin{eqnarray}
W_s^2=\left(\frac{c_0}{c_1}\right)^2B^2\int_{k_{min}}^{k_{max}}p(k)(f_-(k))^2\ dk
+\frac{c_0}{c_1}B\int_{k_{min}}^{k_{max}}p(k) f_+(k)\ dk\ .\nonumber
\end{eqnarray}

The function $f_{\pm}(k)$ has a crossover at $k=k^*$ , where $k^{*}$ is the
crossover degree between the two different behaviors, then

1) for $k<k^{*} \Rightarrow  f_{\pm}(k) \approx \pm \ \langle k \rangle \ \Omega(k)/I_1/2$, and

2) for $k>k^{*} \Rightarrow f_{\pm}(k) \approx 1$.

As $k^{*}$ is the crossover between two different behaviors of $f_{\pm}(k)$,
and the numerator of the function diverges faster than the denominator, we
have $1 \approx \langle k \rangle\ \Omega(k^{*})/I_1$ thus $k^{*} \approx
\ln(I_1/\langle k \rangle)/ \ln (1-c_0)$.  Then,
\begin{eqnarray}\label{rugsatint}
W_s^2&&=\left[\left( \frac{c_0}{c_1} \right)^2
 B^2+\frac{c_0}{c_1}B\right]\int_{k^{*}}^{k_{max}}p(k)dk +\left(\frac{c_0
 B}{2 c_1 I_1}\right)^2\langle k
 \rangle^2\int_{k_{min}}^{k^{*}}p(k)(\Omega(k))^2dk\nonumber \\
 &&+\left(\frac{c_0 B}{2 c_1 I_1}\right)\langle k
 \rangle\int_{k_{min}}^{k^{*}}p(k)\Omega(k)dk\ .\nonumber
\end{eqnarray}

Even thought $k^{*}$ depends on $k_{max}$, it can be demonstrated that the two last
integrals depends weakly on $N$ and can be considerer as constant. Then,
introducing the corrections due to finite size effects trough $k_{max}$ in
$\langle k \rangle$, we obtain
\begin{eqnarray}\label{tamfinito}
W_s^2 \sim W_s^2(\infty)\left( 1+\frac{A_1}{N}+\frac{A_2}{N^{\frac{\lambda -2}{\lambda
-1}}}+\frac{A_3}{N^{2\frac{\lambda -2}{\lambda -1}}}\right)\ .
\end{eqnarray}
where $A_1$, $A_2$ and $A_3$ do not depend on $k_{max}$.

\begin{figure}
\includegraphics[width=0.4\textwidth]{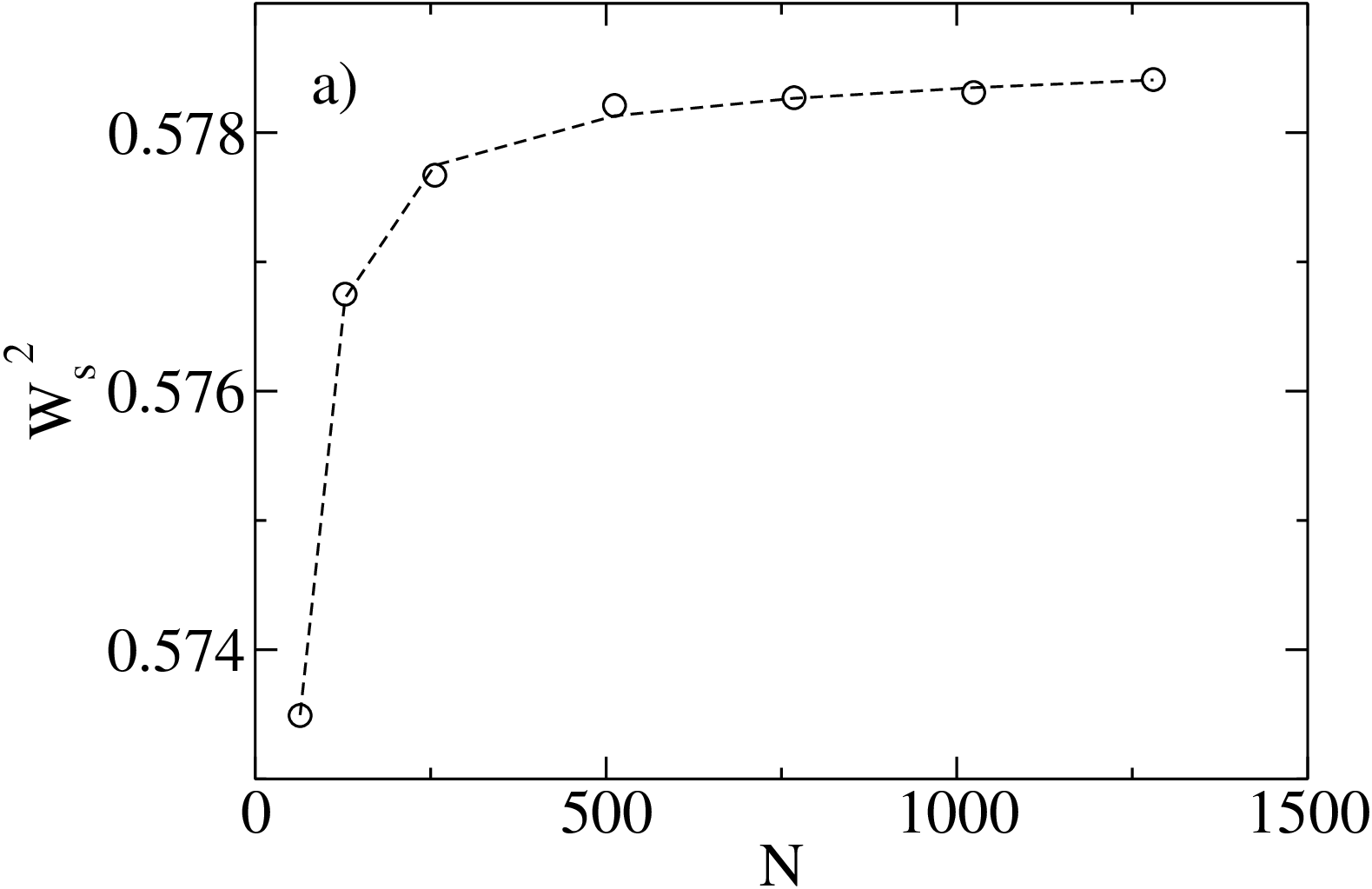}
\hspace{1cm}
\includegraphics[width=0.4\textwidth]{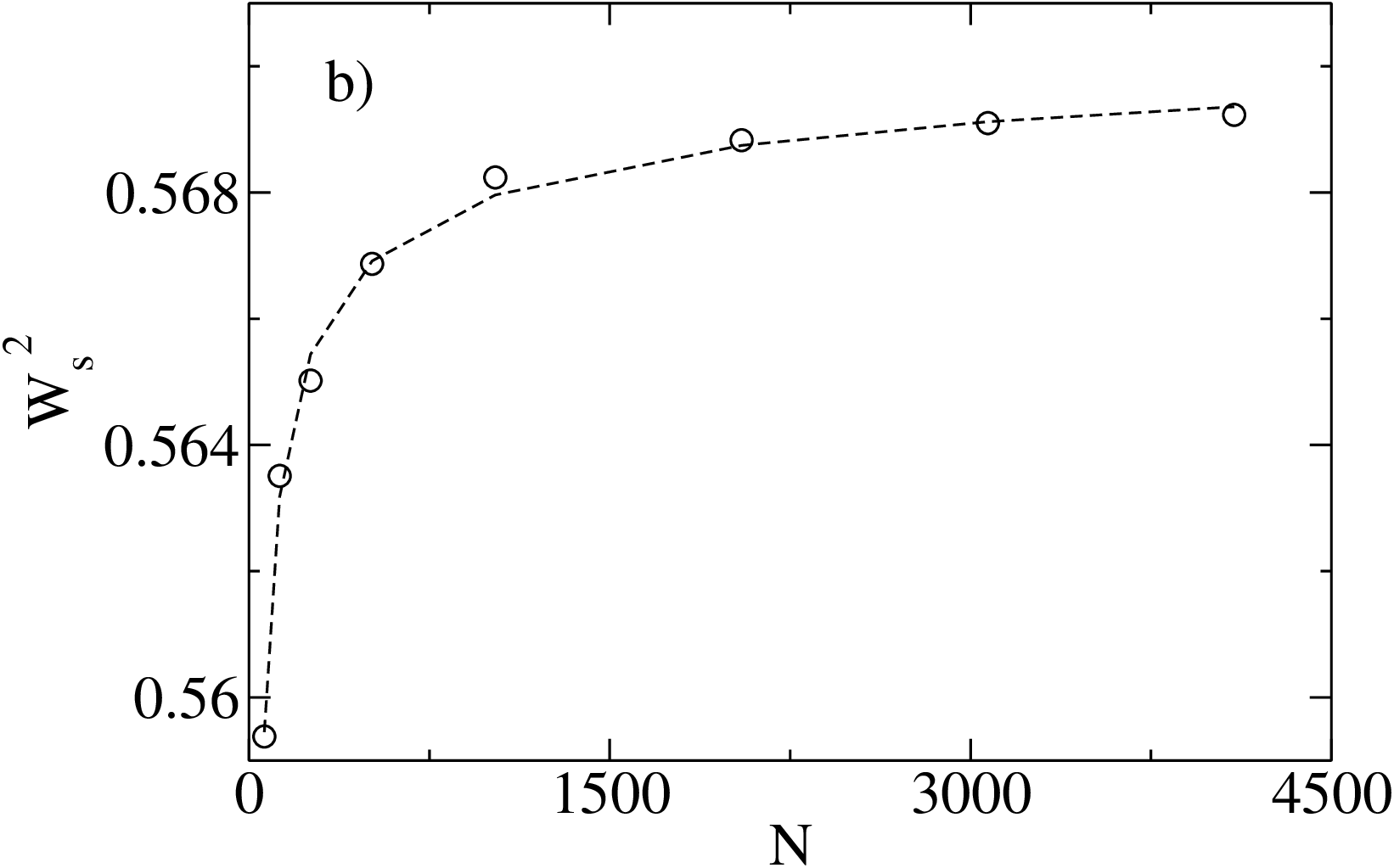}
\caption{$W_s^2$ as a function of $N$ for a) $\lambda=\ 3.5$ and b)
  $\lambda=\ 2.5$ in symbols for the same system sizes of the
  Fig~\ref{fig.1}. The dashed lines represent the fitting with
  Eq.~(\ref{tamfinito}), obtained in the MF approximation by considering the
  finite-size effects introduced by the MR construction.\label{fig.2}}
\end{figure}

\begin{figure}
\includegraphics[width=0.4\textwidth]{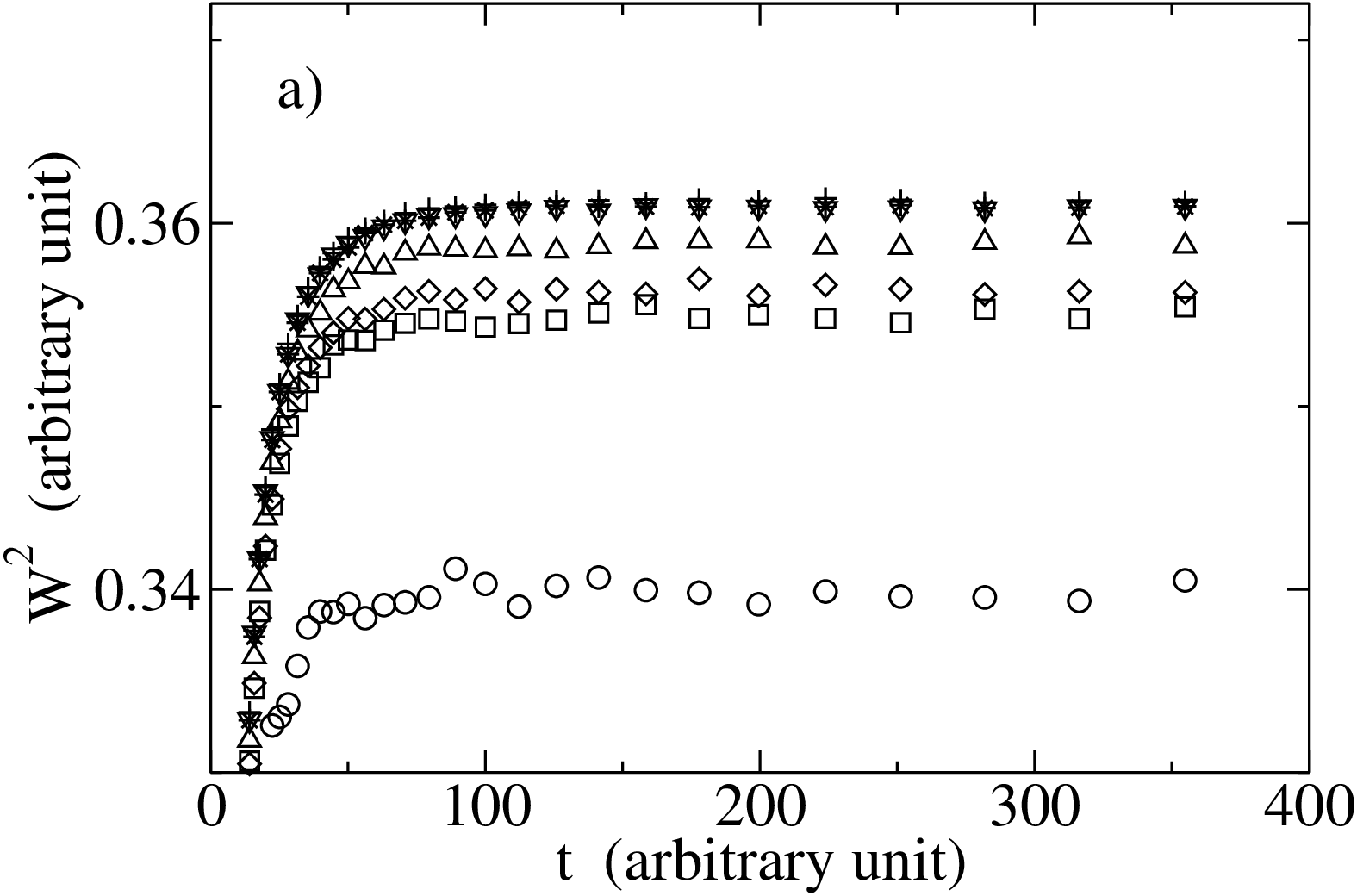}
\hspace{1cm}
\includegraphics[width=0.4\textwidth]{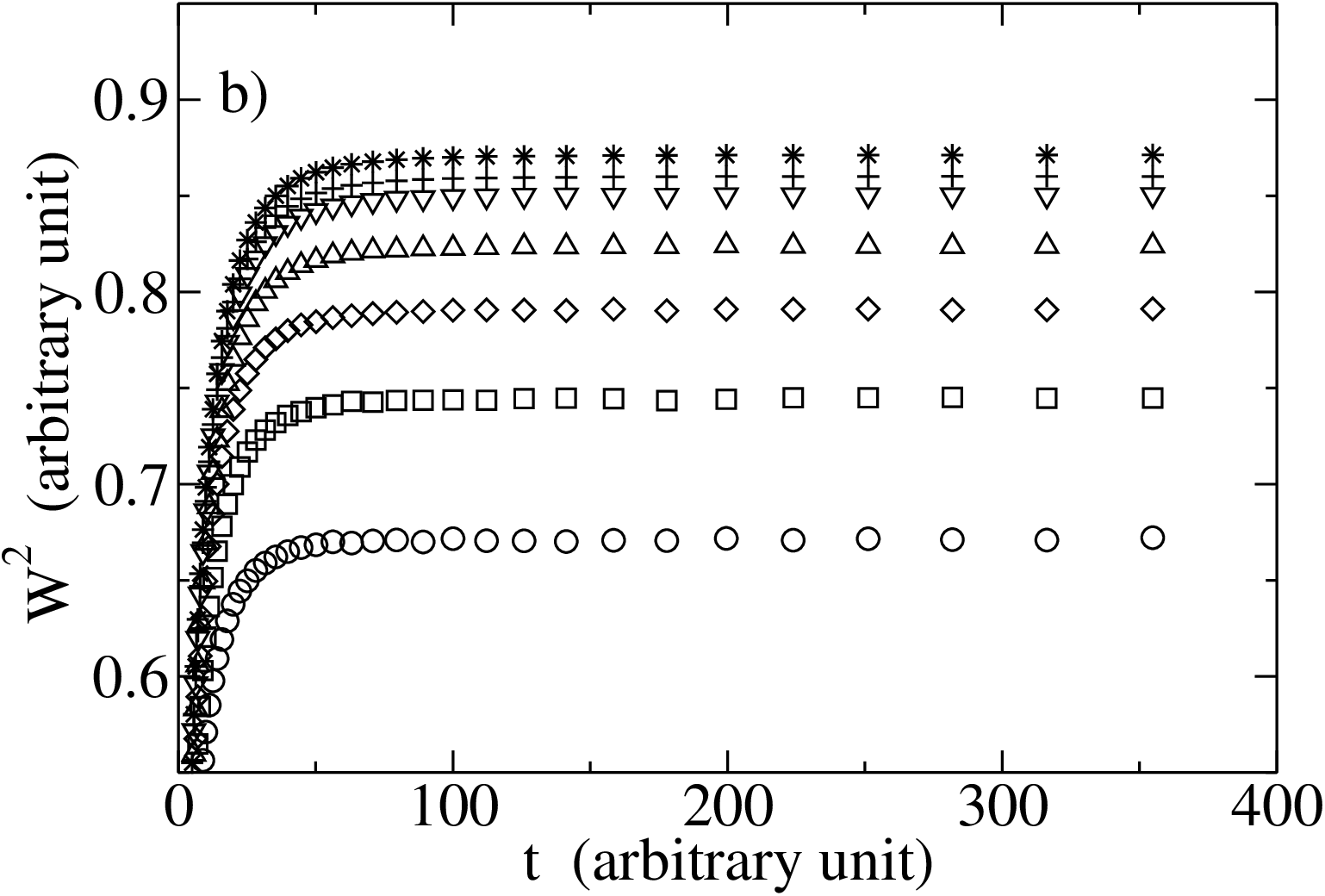}
\caption{$W^2$ as a function of $t$ from the integration of the evolution
 equation: a) $\lambda=\ 3.5$ for $N=64$ ($\circ$), $128$ ($\Box$), $256$
 ($\diamond$), $512$ ($\bigtriangleup$), $1024$ ($\bigtriangledown$), $1536$
 ($+$) and $2048$ ($\star$) and b) $\lambda=\ 2.5$ for $N=64$ ($\circ$),
 $128$ ($\Box$), $256$ ($\diamond$), $512$ ($\bigtriangleup$), $768$
 ($\bigtriangledown$), $1024$ ($+$) and $1280$ ($\star$). For all the
 integrations we use $U= 0.5$ and typically $1.000$ realizations of
 networks.\\\label{fig.3}}
\end{figure}

\begin{figure}
\includegraphics[width=0.4\textwidth]{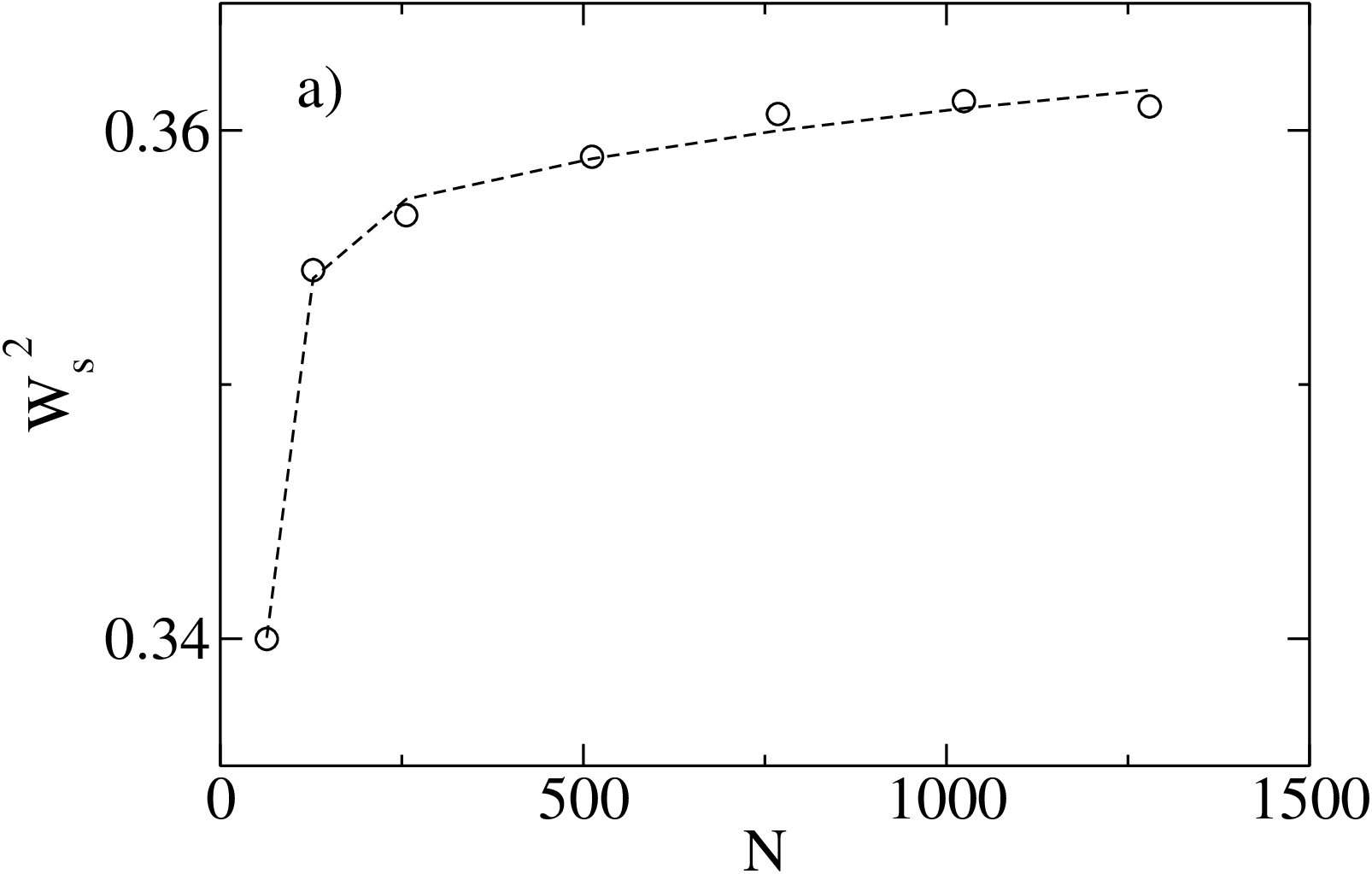}
\hspace{1cm}
\includegraphics[width=0.4\textwidth]{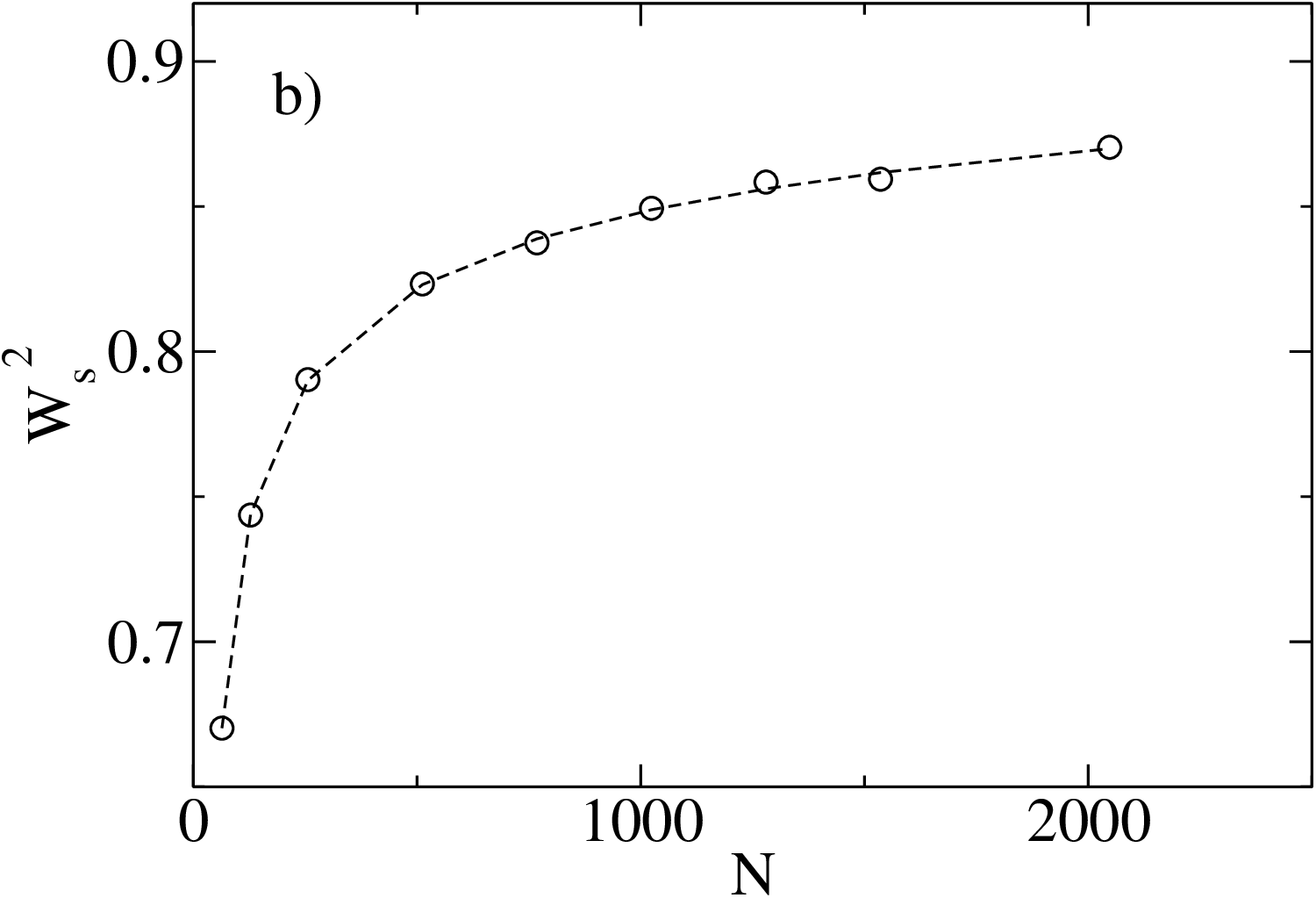}
\caption{$W_s^2$ as a function of $N$ for a) $\lambda=\ 3.5$ and b)
  $\lambda=\ 2.5$ in symbols for the same system sizes of the
  Fig~\ref{fig.3}. The dashed lines represent the fitting with
  Eq.~(\ref{tamfinito}), obtained by considering the finite-size effects
  introduced by the MR construction.\label{fig.4}}
\end{figure}

In Fig.~\ref{fig.2} and Fig.~\ref{fig.4} the dashed lines represent the
fitting of the curves with Eq.~(\ref{tamfinito}) considering finite-size
effects introduced by the MR construction. We can see that this equation
represents very well the finite size effects of this model. This means that
even though the networks is heterogeneous, the non-linear terms are not
necessary to explain the $N$ independence of $W_s$ when a conservative noise
is used. Notice that even when our network is correlated in the degree, the
expression for $W_s^2$ found describe very well the scaling behavior with $N$
as shown in the insets of Fig.~\ref{fig.2} and \ref{fig.4}. This model
suggest a useful load balance algorithm suitable for processors
synchronization in parallel computation. Our results show that the algorithm
could be useful when one want to increase the number of processors and its
general behavior is well represented by a simple mean field equation.

\section{Summary}

In summary, in this paper we study a conservative model in SF networks and
find that the roughness of the steady state is a constant and its dependence
on $N$ for any $\lambda$ it is only due to finite size effects. We derive
analytically the evolution equation for the model, and retain only linear
terms because they are enough to explain the scaling behavior of
$W_s$. Finally, we apply the mean field approximation to the equation and we
calculate explicitly the corrections to scaling of $W_s$. This approximation
describe very well the behavior of the model and shows clearly that the
corrections are due to finite size effects.

\section{Acknowledgments}

This work has been supported by UNMdP and FONCyT (Pict
2005/32353).

\begin{thebibliography}{99}
\bibitem{Motter_prl} A. E. Motter, Phys. Rev. Lett {\bf 93}, 098701 (2004).
\bibitem{Lopez_transport} E. L\'opez {\it et al.}, Phys. Rev. Lett. {\bf
94}, 248701~(2005); A. Barrat, M. Barth\'elemy, R. Pastor-Satorras and
A. Vespignani, PNAS {\bf 101}, 3747~(2004).
\bibitem{zhenhua} Z.~Wu, {\it et al.},  Phys.~Rev.~E.~\textbf{71},
045101(R)~(2005).
\bibitem{Pastorras_PRL_2001} R. Pastor-Satorras and A. Vespignani,
Phys. Rev. Lett. 86, 3200(2001).
\bibitem{Jost-prl} J. Jost and M. P. Joy, Phys. Rev. E {\bf 65}, 016201
(2002); X. F. Wang, Int. J. Bifurcation Chaos Appl. Sci. Eng. {\bf 12}, 885
(2002); M. Barahona and L. M. Pecora, Phys. Rev. Lett. {\bf 89}, 054101
(2002); S. Jalan and R. E. Amritkar, Phys. Rev. Lett. {\bf 90}, 014101
(2003); T. Nishikawa {\it et al.}, Phys. Rev. Lett. {\bf 91}, 014101 (2003);
A. E. Motter {\it et al.}, Europhys. Lett. {\bf 69}, 334 (2005); A. E. Motter
{\it et al.}, Phys. Rev. E {\bf 71}, 016116 (2005).
\bibitem{Korniss07}  G. Korniss, Phys. Rev. E {\bf 75}, 051121 (2007).
\bibitem{Kozma05} H. Guclu, G. Korniss and Z. Toroczkai, Chaos {\bf 17},
026104 (2007).
\bibitem{Nagurney} A. Nagurney, J. Cruz, J. Dong, and D. Zhang,
Eur. J. Oper. Res. {\bf 164}, 120 (2005).
\bibitem{JWScanell} J. W. Scannell {\it et al.}, Cereb. Cortex 9, 277 (1999).
\bibitem{eubank_2004} S. Eubank, H. Guclu, V. S. A. Kumar, M. Marathe,
A. Srinivasan, Z.  Toroczkai and N. Wang, Nature {\bf 429}, 180 (2004);
M. Kuperman and G. Abramson, Phys Rev Lett {\bf 86}, 2909 (2001).
\bibitem{anita} A. L. Pastore y Piontti, P. A. Macri and L. A. Braunstein,
  Phys. Rev. E {\bf 76}, 046117 (2007).
\bibitem{Barabasi_sf} R. Albert and A.-L. Barab\'{a}si, Rev. Mod. Phys.
 {\bf 74}, 47 (2002); S. Boccaletti, V. Latora, Y. Moreno, M. Chavez and
 D.-U. Hwang, Physics Report {\bf 424}, 175 (2006).
\bibitem{family} F. Family, J Phys. A {\bf 19}, L441 (1986).
\bibitem{crispre} C. E. La Rocca, L. A. Braunstein and P. A. Macri,
Phys. Rev. E {\bf 77}, 046120 (2008).
\bibitem{cneucli1} Youngkyun Jung and In-mook Kim, Phys. Rev. E {\bf 59}, 7224 (1999).
\bibitem{cneucli2} In-mook Kim, Jin Yang and Youngkyun Jung, Journal of the
Korean Physical Society {\bf 34}, 314 (1999).
\bibitem{EW} S. F. Edwards and D. R. Wilkinson, Proc. R. Soc. London, Ser. A
  {\bf 381}, 17 (1982).
\bibitem{Molloy} M. Molloy and B. Reed, Random Struct. Algorithms
{\bf 6}, 161 (1995); Combinatorics, Probab. Comput. {\bf 7}, 295 (1998).
\bibitem{boguna} M. Bogu\~n\'a, R. Pastor-Satorras, and A. Vespignani,
Eur. Phys. J. B {\bf 38}, 205 (2004).
\bibitem{cohen} R. Cohen, S. Havlin, and D. ben-Avraham 446. Chap. 4 in
  "Handbook of graphs and networks", Eds. S. Bornholdt and H. G. Schuster,
  (Wiley-VCH, 2002).
\bibitem{VK} N. G. Van Kampen, {\it Stochastic Processes in Physics and
Chemistry}, North-Holland, Amsterdam (1981).
\bibitem{Vveden} D. D. Vvedensky, Phys. Rev. E {\bf 67}, 025102(R) (2003).
\bibitem{lidia} L. A. Braunstein, R. C. Buceta, C. D. Archubi and
G. Costanza, Phys. Rev. E {\bf 62}, 3920 (2000).
\bibitem{ruidoconser} A. Ballestad, B. J. Ruck, J. H. Schmid, M. Adamcyk,
E. Nodwell, C. Nicoll, and T. Tiedje, Phys. Rev. B {\bf 65}, 205302 (2004).
\bibitem{pasointeg} B. Kozma, M. B. Hastings and G. Korniss,
J. Stat. Mech. Theor. Exp. (2007) P08014.
\end{thebibliography}
\end{document}